\documentclass[twocolumn,showpacs,pra]{revtex4}

\usepackage{graphicx}

\begin{document}

\title{Analytic Evaluation of Four-Particle Integrals\\
with Complex Parameters}
\author{V.~S.~Zotev}
\affiliation{Department of Physics, University of California,
Riverside, California 92521}
\author{T.~K.~Rebane}
\affiliation{Institute of Physics, St.~Petersburg State
University, ul.~Ulyanovskaya 1, Petrodvorets, 198904 Russia}
\date{March 4, 2002}

\begin{abstract}
The method for analytic evaluation of four-particle integrals,
proposed by Fromm and Hill, is generalized to include complex
exponential parameters. An original procedure of numerical branch
tracking for multiple valued functions is developed. It allows
high precision variational solution of the Coulomb four-body
problem in a basis of exponential-trigonometric functions of
interparticle separations. Numerical results demonstrate high
efficiency and versatility of the new method.
\end{abstract}


\maketitle

\section{Introduction}

The problem of four particles with the Coulomb interaction plays
an important role in atomic and nuclear physics. It forms a link
between the three-body problem, which can be solved numerically
with very high precision, and many-body problems, solutions of
which are very approximate. Thus, profound studies of various
four-particle systems can provide valuable insights into physics
of systems with greater numbers of particles.

In addition to the methodological interest, the four-body problem
has unquestionable practical significance. Positronium beams are
extensively used in positronium-atom scattering experiments, but
the positronium molecule, $e^{+}e^{-}e^{+}e^{-}$, has not been
observed experimentally yet. All existing knowledge of its
properties is based on numerical studies \cite{usu98}. Molecules
and ions including $\mu$-meson have attracted much attention
traditionally in connection with the problem of the muon catalyzed
fusion. Calculations suggest \cite{fro97} that muonic molecules
like $p^{+}\mu^{-}p^{+}\mu^{-}$ have higher nuclear reaction rates
than the corresponding three-particle ions. These examples show
that high-precision numerical solution of the four-body problem is
essential for proper understanding of various physical phenomena.

The majority of four-particle systems are nonadiabatic, and cannot
be treated within the adiabatic approximation. The only practical
way to calculate their energy and properties is to use the
variational approach, taking into account the correlated motion of
all the particles. Basis functions of the Gaussian type, depending
on six interparticle separations and several nonlinear parameters,
have been extensively used for such calculations
\cite{usu98,fro97,suz98}. An important advantage of the Gaussian
functions is that all integrals can be easily evaluated. The
nonlinear parameters are optimized stochastically \cite{suz98}: at
each step of basis expansion, many functions with randomly
generated parameters are examined, and the function, giving the
largest decrease in energy, is added to the basis.

However, unlike real wavefunctions, the Gaussian functions do not
decay exponentially, and do not satisfy the cusp condition. From
this point of view, they are rather unphysical. As a result,
convergence of the variational procedure is very slow, and many
hundreds of basis functions must be used. A recent calculation of
the positronium molecule \cite{usu98} involved 1600 Gaussian
functions. It was suggested that further expansion of the basis
was not practical because of increasing computation time and low
probability of finding good parameters. Thus, more efficient basis
functions are clearly required.

A method for analytic evaluation of four-particle integrals,
proposed by Fromm and Hill \cite{hil87}, opened up possibilities
of variational calculation of four-particle systems in a basis of
exponential functions of interparticle separations. This method
reduces computation of integrals, needed to determine matrix
elements of a four-particle Hamiltonian, to evaluation of the
dilogarithmic function \cite{lev81} of various arguments.
Application of this method, however, is a very difficult problem.
Because the dilogarithm is a multiple valued function, the entire
algorithm cannot be used without an effective procedure of branch
and singularity tracking.

This problem was initially solved by the authors for the case of
real exponential parameters. The first calculations of the
positronium molecule \cite{reb96}, and several mesic molecules
\cite{zot98} in the exponential basis, depending on all six
interparticle separations, have demonstrated high efficiency and
great potential of this method. To the best of our knowledge,
nobody else has done this yet \cite{fro01}.

Because one exponential function is as effective as eight
Gaussians, a size of the basis can be reduced significantly.
However, an amount of time, needed to compute one matrix element,
is much larger than for the Gaussian basis. Thus, optimization of
nonlinear parameters is the main difficulty. Deterministic
optimization (gradient descent) gives excellent results for a
relatively small number of exponential basis functions. Stochastic
optimization (trial and error), used to expand the basis further,
is inefficient due to a dramatic increase in computation time.
This fact suggests that a possible alternative to an enormously
large Gaussian basis is a relatively short basis of the most
efficient and versatile functions with carefully optimized
parameters.

A natural generalization of the exponential basis is the
exponential-trigonometric basis, obtained by replacing real
exponential parameters with complex ones \cite{reb90}. The
exponential-trigonometric functions have been successfully
employed in variational calculations of three-particle adiabatic
systems \cite{zot94}. They are much more efficient, than the
ordinary exponentials for two reasons. First, they contain twice
as many nonlinear parameters, thus allowing better approximation
of the wavefunction. Second, they exhibit nonmonotonic dependence
on interparticle separations, being able to imitate sharp peaks in
wavefunctions of adiabatic systems. The computation time increases
only insignificantly in comparison with the case of real
exponential parameters.

In order to use the exponential-trigonometric basis in the
four-body problem, one has to evaluate the four-particle integrals
with complex parameters. The problem of branch tracking in a
general complex case is formidable. Every branch change for every
multiple valued function has to be taken into account if correct
values of the integrals are to be obtained. An original (and,
inevitably, very nontrivial) procedure of numerical branch
tracking has been developed by the authors. The first variational
calculations of four-particle systems in the
exponential-trigonometric basis proved extremely promising
\cite{zot00}. They showed that one exponential-trigonometric
function can replace seven exponential functions in calculation of
$e^{+}e^{-}e^{+}e^{-}$, and several dozen exponentials in studies
of adiabatic systems \cite{zot00}. Therefore, it presents a real
alternative to both the exponential and Gaussian basis functions.

Even though results of the calculations involving the exponential
and exponential-trigonometric functions have been published
\cite{reb96,zot98,zot00}, details of the new method have not been
reported yet. The purpose of the present paper is to fill this
gap. We present a description of our algorithm that will enable a
reader to implement it as a computer program.

The paper is organized as follows. Section~II.A discusses what
integrals are needed to compute matrix elements of a four-particle
Hamiltonian, and how the number of them can be reduced. In
Sec.~II.B, principles of the original method by Fromm and Hill are
outlined. Sec.~II.C provides information about multiple valued
functions used in the analysis. In Sec.~II.D, a simplified
procedure of branch tracking in the case of real parameters is
described. Sec.~II.E gives a detailed exposition of the method of
branch tracking in the most general case, when all the parameters
are complex. In Sec.~II.F, a practical implementation of the
branch tracking algorithm is described. The last Section presents
our conclusions.

\section{Description of the method}

\subsection{Matrix elements of four-particle Hamiltonian}

Let us consider a Hamiltonian of a four-particle system with the
Coulomb interactions:
\begin{equation}
H=-\frac{\hbar^{2}}{2} \sum_{j=1}^{4} \frac{\Delta_{j}}{m_{j}} +
\sum_{j<k}^{4} \frac{q_{j}q_{k}}{r_{jk}}~~.
\end{equation}
Here $m_{j}$ and $q_{j}$, $j=1...4$, are masses and charges, and
$r_{jk}=|\mathbf{r}_{j}-\mathbf{r}_{k}|$ are interparticle
separations. Our purpose is to evaluate matrix elements of $H$
with exponential basis functions
\begin{equation}
\Phi_{b}=\exp ( -\sum_{j<k}^{4}b_{jk}r_{jk} ),~~~~
\Phi_{c}=\exp ( -\sum_{j<k}^{4}c_{jk}r_{jk} ).
\end{equation}
These functions depend on complex parameters $\{b_{jk}\}$ and
$\{c_{jk}\}$. In what follows, the notation $\{x_{jk}\}$ will
always refer to six quantities, $x_{jk}$, with $j,k=1...4$ and
$j<k$, i.e. $x_{12},x_{13},x_{14},x_{23},x_{24},x_{34}$, assuming
that $x_{jk}=x_{kj}$.

In order to compute matrix elements of the operator of kinetic
energy in Eq.~(1), one has to evaluate the following quantities:
\begin{equation}
\langle \Phi_{b} \mid \cos \Theta_{jkl} \mid \Phi_{c} \rangle =
\langle \Phi_{b} \mid
\frac{r_{jk}^{2}+r_{jl}^{2}-r_{kl}^{2}}{2r_{jk}r_{jl}} \mid
\Phi_{c} \rangle~,
\end{equation}
where $j \neq k$, $k \neq l$, $j \neq l$. The integrands in the
last formula display linear and even quadratic dependence on
certain interparticle separations. Therefore, in order to obtain
matrix elements of the Hamiltonian, Eq.~(1), one has to calculate
a total of 43 integrals: one overlap integral, six integrals of
the Coulomb interactions, and 36 integrals, given by Eq.~(3).

It turns out, however, that it is possible to avoid computation of
the integrals in Eq.~(3). It has been shown by one of the authors
that the matrix elements of the above Hamiltonian can be expressed
in terms of the overlap integral and six Coulomb integrals only
\cite{reb93}. Thus, one can write:
\begin{equation}
\langle \Phi_{b} \mid H \mid \Phi_{c} \rangle =
H_{1}-H_{2}-H_{3}~~.
\end{equation}
The individual terms in Eq.~(4) are given by the following
expressions \cite{reb93}:
\begin{equation}
H_{1}=\sum_{j<k}^{4} \left[  \frac{(m_{j}+m_{k})}{2m_{j}m_{k}}
a_{jk}+q_{j}q_{k} \right] \langle \Phi_{a} \mid \frac{1}{r_{jk}}
\mid \Phi_{a} \rangle~~;
\end{equation}
\begin{equation}
H_{2}=\sum_{j<k}^{4} \frac{(m_{j}+m_{k})}{2m_{j}m_{k}}~
d_{jk}^{\,2}~ \langle \Phi_{a} \mid \Phi_{a} \rangle~~;
\end{equation}
\begin{eqnarray}
H_{3}=\sum_{j=1}^{4}\sum_{\renewcommand{\arraystretch}{0.7}
\begin{array}{c} \scriptstyle k<l
\\ \scriptstyle k,l \neq j \end{array}}^{4}
\frac{(a_{jk}s_{jk}+a_{jl}s_{jl}-a_{jn}s_{jn})}{2m_{j}\,a_{jk}a_{jl}}~
d_{jk}d_{jl}~.
\end{eqnarray}

In these formulas, $\Phi_{a}$ is a new function with parameters
$\{a_{jk}\}$, defined as $a_{jk}=(b_{jk}+c_{jk})/2$:
\begin{equation}
\Phi_{a}=\exp(-\sum_{j<k}^{4}a_{jk}r_{jk})~~.
\end{equation}
The parameters $\{d_{jk}\}$ are defined as
$d_{jk}=(c_{jk}-b_{jk})/2$, and the quantities $\{s_{jk}\}$ are
given by:
\begin{equation}
s_{jk}=\langle \Phi_{a} \mid \frac{1}{r_{jk}} \mid \Phi_{a}
\rangle - a_{jk} \langle \Phi_{a} \mid \Phi_{a} \rangle~~.
\end{equation}
The additional index $n$ in Eq.~(7) is fixed by a condition $n
\neq j,k,l$.

Therefore, only seven integrals -- the overlap integral and six
Coulomb integrals, calculated with the function $\Phi_{a}$, -- are
needed to determine the matrix elements of the Hamiltonian,
Eq.~(1). The above formulas are valid for both real and complex
parameters. They are indispensable for any application of this
method.

\subsection{Evaluation of four-particle integrals}

In this paper, we generalize the method of analytic evaluation of
four-particle integrals, proposed by Fromm and Hill \cite{hil87},
to include complex exponential parameters. First, we would like to
recall basic ideas of this method. The following family of
integrals is considered:
\begin{equation}
J(\{n_{jk}\},\{\alpha_{jk}\})=\int ( \prod_{j<k}^{4}
r_{jk}^{n_{jk}-1} ) \exp(-\sum_{j<k}^{4} \alpha_{jk}r_{jk})~dV~~.
\end{equation}
Here, $\{\alpha_{jk}\}$ denotes a set of six exponential
parameters,
$\alpha_{12},~\alpha_{13},~\alpha_{14},~\alpha_{23},~\alpha_{24},~\alpha_{34}$,
and $\{n_{jk}\}$ is the corresponding set of non-negative
integers. The integrand depends on six interparticle separations,
$\{r_{jk}\}$. The integration is performed over 9-dimensional
space of relative coordinates of four particles:
$dV=d^{\,3}r_{12}d^{\,3}r_{13}d^{\,3}r_{14}$.

An integral with all $n_{jk}=0$ is called ``generating'':
\begin{equation}
I(\{\alpha_{jk}\})=\int ( \prod_{j<k}^{4} r_{jk}^{-1} )
\exp(-\sum_{j<k}^{4} \alpha_{jk}r_{jk})~dV~~.
\end{equation}
All the integrals in Eq.~(10) can be obtained from the generating
integral, Eq.~(11), by differentiation:
\begin{equation}
J(\{n_{jk}\},\{\alpha_{jk}\})=[
\prod_{j<k}^{4}(-\frac{\partial}{\partial \alpha_{jk}})^{n_{jk}} ]
\, I(\{\alpha_{jk}\})~~.
\end{equation}
The generating integral is given by the following formula:
\begin{equation}
I(\{\alpha_{jk}\})=\frac{16\pi^{3}}{\sigma} [ \sum_{j=1}^{4}
\sum_{k=1}^{4} v(\gamma_{k}^{(j)} / \sigma) + \sum_{j=2}^{4}
u(\beta_{1}^{(1)}\beta_{1}^{(j)}) ]~~.
\end{equation}
The functions $v(z)$ and $u(z)$ are expressed in terms of the
dilogarithmic function $\mathrm{Li}_{2}(z)$:
\begin{equation}
u(z)=\mathrm{Li}_{2}(z)-\mathrm{Li}_{2}(1/z)~~;
\end{equation}
\begin{eqnarray}
v(z)=\frac{1}{2}~\mathrm{Li}_{2}[(1-z)/2]-\frac{1}{2}~\mathrm{Li}_{2}[(1+z)/2]-
\\ -\frac{1}{4}\ln^{2}[(1-z)/2]+\frac{1}{4}\ln^{2}[(1+z)/2]~~.
\nonumber
\end{eqnarray}

In Eq.~(13) for the generating integral, $\gamma_{k}^{(j)}$ are
third-order polynomials in $\alpha$'s, defined in the following
way:
\begin{eqnarray}
\gamma_{k}^{(j)}=-\mu_{j}^{(j)}-\mu_{k}^{(j)}+\mu_{l}^{(j)}+\mu_{m}^{(j)}~~,\\
\gamma_{j}^{(j)}=+\mu_{1}^{(j)}+\mu_{2}^{(j)}+\mu_{3}^{(j)}+\mu_{4}^{(j)}~~,
\nonumber
\end{eqnarray}
where for each $j \neq k$: $l \neq j,k$; $m \neq j,k$; $l \neq m$.
The polynomials $\mu_{k}^{(j)}$ are defined as follows:
\begin{eqnarray}
\mu_{k}^{(j)}=\alpha_{lm}(-\alpha_{jk}^{2}+\alpha_{kl}^{2}+\alpha_{km}^{2})~~,~\\
\mu_{j}^{(j)}=2\alpha_{lm}\alpha_{kl}\alpha_{km}~~,~~~~~~~~~~~~~~~~
\nonumber
\end{eqnarray}
with the same restrictions on values of $j,~k,~l,$ and $m$.

The function $\sigma$ is a square root of a sixth-order polynomial
in $\alpha$'s: $\sigma=\sqrt{s_{1}+s_{2}}$. The quantity $s_{1}$
in this expression is given by
\begin{equation}
s_{1}=\sum_{j=2}^{4}
\alpha_{1j}^{2}\alpha_{lm}^{2}(\alpha_{1j}^2+\alpha_{lm}^{2}-\alpha_{1l}^2-
\alpha_{1m}^{2}-\alpha_{jl}^2-\alpha_{jm}^{2})~~,
\end{equation}
where for each $j$: $l \neq 1,j$; $m \neq 1,j$; $l \neq m$. The
quantity $s_{2}$ is determined as
\begin{equation}
s_{2}=\sum_{j=1}^{4}
\alpha_{jl}^{2}\alpha_{jm}^{2}\alpha_{jk}^{2}~~,
\end{equation}
where for each $j$: $l,m,k \neq j$; $l \neq m$; $m \neq k$; $l
\neq k$. Finally, $\beta_{k}^{(j)}$ is defined by the following
expression:
\begin{equation}
\beta_{k}^{(j)}=(\sigma - \gamma_{k}^{(j)}) / (\sigma +
\gamma_{k}^{(j)})~~.
\end{equation}

In all these formulas, indices $j,k,l,m$ change from 1 to 4, and
it is assumed that $\alpha_{jk}=\alpha_{kj}$ for each $j \neq k$.
If some indices are not defined uniquely, the formulas are
symmetric under their permutations.

Eq.~(13) is the main result of this method \cite{hil87}. It
provides an analytic expression for the generating integral,
Eq.~(11). It was pointed out \cite{hil87} that there is no need to
know an analytic dependence of the generating integral on the
parameters $\{\alpha_{jk}\}$ to compute the family of integrals,
Eq.~(10). According to Eq.~(12), all these integrals are
derivatives of the generating integral. Special formulas can be
used \cite{hil87} to calculate numerical values of derivatives of
functions $f \cdot g$ and $h(g)$, if numerical values of
derivatives of the functions $f,~g$, and $h$ have already been
computed. For example, derivatives of the term
$v(\gamma_{k}^{(j)}/\sigma)$ in Eq.~(13) can be obtained in the
following way. First, derivatives of $\sigma^{2}$ and
$\gamma_{k}^{(j)}$ with respect to $\{\alpha_{jk}\}$ are
calculated. Then derivatives of a function $h(z)=z^{-1/2}$ are
computed at $z=\sigma^{2}$. After that, using a formula for
derivatives of $h(g)$ with $g=\sigma^{2}$, one finds derivatives
of $1/\sigma$ with respect to $\{\alpha_{jk}\}$. Then, using a
formula for derivatives of $f \cdot g$ with $f=\gamma_{k}^{(j)}$
and $g=1/\sigma$, one obtains derivatives of $\gamma_{k}^{(j)}$
with respect to $\{\alpha_{jk}\}$. After that, derivatives of a
function $h(z)=v(z)$ at $z=\gamma_{k}^{(j)}/\sigma$ are
calculated. Finally, using a formula for derivatives of $h(g)$
with $g=\gamma_{k}^{(j)}/\sigma$, one can find the derivatives of
$v(\gamma_{k}^{(j)}/\sigma)$ with respect to $\{\alpha_{jk}\}$.
Within this approach, all the integrals of Eq.~(10) can be
evaluated by means of an efficient recursive procedure, working
with numbers only.

At this point, we can appreciate importance of Eqs.~(4)-(9). In
order to obtain the matrix elements of the Hamiltonian, Eq.~(1),
we have to compute the mixed derivatives given by Eq.~(12) up to
the sixth order only, i.e. for $n_{jk}=0,1$, where $j,k=1...4$,
and $j<k$. This means that, at each step of the recursive
procedure, we calculate $2^{6}=64$ derivatives. If we tried to
evaluate the integrals of Eq.~(3) directly, it would be necessary
to compute the mixed derivatives in Eq.~(12) up to 18th order,
i.e. for $n_{jk}=0...3$. The number of derivatives, calculated at
each step, would increase quadratically. An amount of time,
required to carry out the entire recursive procedure, would be
enormous. Therefore, the original method by Fromm and Hill
\cite{hil87}, used by itself, does not make high precision
calculations of four-particle systems possible. Only in
conjunction with the method \cite{reb93} for reducing the number
of integrals can it produce valuable results.

\subsection{The multiple valued functions}

The main difficulty in using Eq.~(13) for the generating integral
is the fact that the functions in this formula are multiple
valued. Indeed, the functions $u(z)$ and $v(z)$, given by
Eqs.~(14) and (15), are expressed in terms of the dilogarithmic
function $\mathrm{Li}_{2}(z)$. The dilogarithm is defined as
follows \cite{lev81}:
\begin{equation}
\mathrm{Li}_{2}(z)=-\int_{0}^{z} \frac{\ln(1-\zeta)}{\zeta}
d\zeta~~.
\end{equation}
This function is analytic inside the unit circle in the complex
plane:
\begin{equation}
\mathrm{Li}_{2}(z)=\sum_{n=1}^{\infty} \frac{z^{n}}{n^{2}}~,~~~~
|z| < 1~~.
\end{equation}
Its values outside the unit circle can be determined using a
relation \cite{lev81}:
\begin{equation}
\mathrm{Li}_{2}(z)=-\frac{\pi^{2}}{6}-\frac{1}{2}\ln^{2}(-z)-
\mathrm{Li}_{2}(1/z)~~.
\end{equation}
In the immediate vicinity of the unit circle, where convergence of
the series in Eq.~(22) is slow, the following relations can be
used to shift the argument of $\mathrm{Li}_{2}(z)$:
\begin{equation}
\mathrm{Li}_{2}(z)=\frac{\pi^{2}}{6}-\ln(z)\ln(1-z)-
\mathrm{Li}_{2}(1-z)~~.
\end{equation}
\begin{equation}
\mathrm{Li}_{2}(z)=\frac{1}{2} \,
\mathrm{Li}_{2}(z^{2})-\mathrm{Li}_{2}(-z)~~.
\end{equation}

Presence of the logarithm in Eqs.~(23) and (24) clearly indicates
that the function $\mathrm{Li}_{2}(z)$ is, in general, multiple
valued. In order to specify its principal branch we need to fix
the principal branch of the logarithm. The complex logarithm has
branch points at 0 and $\infty$. We choose its branch cut to run
along the negative real axis and define the principal branch as
follows:
\begin{equation}
\ln(z)=\ln|z|+i\,\mathrm{arg}\,z~,~~~~-\pi<\mathrm{arg}\,z<\pi~~.
\end{equation}
This choice determines branch cuts and fixes the principal branch
for the dilogarithm, and the functions $u(z)$ and $v(z)$.

The function $\mathrm{Li}_{2}(z)$ has branch points at 1 and
$\infty$; its branch cut runs from 1 to $\infty$ along the
positive real axis. The function $u(z)$ has branch points at 0, 1,
and $\infty$; its branch cut goes from 0 to $\infty$ along the
positive real axis. The function $v(z)$ has branch points at 1,
-1, and $\infty$; its branch cuts run from $\infty$ to -1 along
the negative real axis, and from 1 to $\infty$ along the positive
real axis.

Fig.~1 exhibits the branch points and cuts for these multiple
valued functions.

\begin{figure}
\resizebox{0.9\columnwidth}{!}{\includegraphics{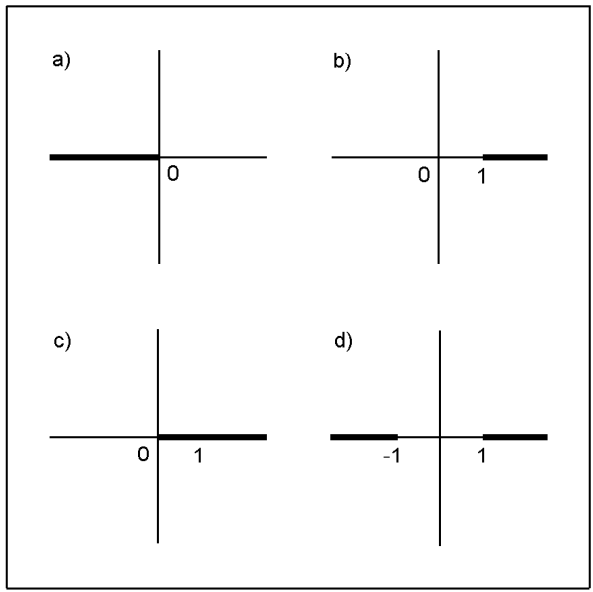}}
\caption{Branch cuts in the complex $z$ plane, necessary to define
principal branches of the multiple valued functions: a) $\ln(z)$,
branch points at 0 and $\infty$; b) $\mathrm{Li}_{2}(z)$, branch
points at 1 and $\infty$; c) $u(z)$, branch points at 0, 1, and
$\infty$; d) $v(z)$, branch points at 1, -1, and $\infty$.}
\end{figure}

It is important to note that the function $\sigma$, which is
present in Eq.~(13) and defined using Eqs.~(18) and (19), is also
a multiple valued function. The complex square root has branch
points at 0 and $\infty$. We choose its branch cut to run along
the positive real axis and define the principal branch as follows:
\begin{equation}
\sqrt{z}=\sqrt{|z|}\exp(\frac{i}{2}\mathrm{arg}\,z)~,~~~~
0<\mathrm{arg}\,z<2\pi~~.
\end{equation}

It can be seen from the definition of the generating integral,
Eq.~(11), that it is a continuous function of parameters
$\{\alpha_{jk}\}$ for all values of these parameters satisfying
the following conditions:
\begin{eqnarray}
~\alpha_{12}+\alpha_{13}+\alpha_{14}>0~, &
~\alpha_{12}+\alpha_{23}+\alpha_{24}>0~, \\
~\alpha_{13}+\alpha_{23}+\alpha_{34}>0~, &
~\alpha_{14}+\alpha_{24}+\alpha_{34}>0~. \nonumber
\end{eqnarray}
\begin{eqnarray}
\alpha_{12}+\alpha_{13}+\alpha_{24}+\alpha_{34}>0~, \nonumber \\
\alpha_{12}+\alpha_{14}+\alpha_{23}+\alpha_{34}>0~, \\
\alpha_{13}+\alpha_{14}+\alpha_{23}+\alpha_{24}>0~.\, \nonumber
\end{eqnarray}
These conditions mean, physically, that the wavefunction of a
system of four particles decreases exponentially when any of the
interparticle separations become infinitely large. If the
parameters $\{\alpha_{jk}\}$ are complex, the above inequalities
must be satisfied by their real parts.

The continuity of the generating integral, Eq.~(11), implies that
the right hand side of Eq.~(13) is also a continuous function of
$\{\alpha_{jk}\}$. This fact has two important consequences.

First, the multiple valued functions $u(z)$, $v(z)$, and
$\sigma(z)$ in Eq.~(13) remain continuous while their branches
change. As a point of interest moves in 12-dimensional space of
six complex parameters, $\{\alpha_{jk}\}$, the arguments of these
functions move freely in the complex plane, and their branches
change repeatedly. However, a computer can evaluate only the
principal branch of the logarithm, given by Eq.~(26), and the
principal branch of the square root, given by Eq.~(27). Therefore,
only the principal branches of the functions $\mathrm{Li}_{2}(z)$,
$u(z)$, $v(z)$, and $\sigma(z)$, defined in the complex plane with
the branch cuts, can be calculated directly. Thus, a special
procedure of branch tracking is necessary to restore continuity of
these functions every time their arguments cross the branch cuts.

Second, all singularities, which different terms in Eq.~(13) can
have, cancel mutually. These singularities arise when $\sigma=0$,
and when any of the following equalities are satisfied:
\begin{eqnarray}
-\alpha_{jl}+\alpha_{jm}+\alpha_{jn}=0~, \nonumber \\
~\alpha_{jl}-\alpha_{jm}+\alpha_{jn}=0~, \\
~\alpha_{jl}+\alpha_{jm}-\alpha_{jn}=0~,\, \nonumber
\end{eqnarray}
where for each $j=1...4$: $l,m,n \neq j$; $l \neq m$; $m \neq n$;
$l \neq n$. These singularities are unphysical, and should have no
effect on the value of the generating integral. As a point under
consideration moves in the space of the parameters
$\{\alpha_{jk}\}$, the arguments of the functions $u(z)$ and
$v(z)$ can frequently appear in the vicinity of the singular
(branch) points. As a result, the values of these functions can
exhibit considerable change, even if the parameters
$\{\alpha_{jk}\}$ change only slightly. Therefore, a special
procedure for dealing with the singularities is needed in order to
carry out explicit cancellation of all singular terms.

This discussion demonstrates that the method of \cite{hil87} is
impossible to use without an effective algorithm for numerical
branch and singularity tracking.

\subsection{Branch tracking in the real case}

Before discussing a general algorithm of branch tracking, it is
beneficial to consider a particular case, when all the exponential
parameters, $\{\alpha_{jk}\}$, are real numbers. Let us introduce
the following parametrization:
\begin{equation}
\alpha_{jk}(p)=(\alpha_{jk}-1)p+1~,~~~~0 \leq p \leq 1~~.
\end{equation}
As the real parameter $p$ changes from 0 to 1, the corresponding
point in 6-dimensional space moves from (1,1,1,1,1,1) to
$(\alpha_{12},\alpha_{13},\alpha_{14},\alpha_{23},\alpha_{24},\alpha_{34})$.
If the parameters $\{\alpha_{jk}\}$ satisfy the conditions of
Eqs.~(28) and (29), the parameters $\{\alpha_{jk}(p)\}$ will
satisfy these conditions for any $p$ between 0 and 1. Therefore,
the generating integral, given by Eq.~(13), must be a continuous
function of $p$. It is known \cite{hil87} that Eq.~(13) with the
functions $u(z)$, $v(z)$, and $\sigma(z)$, represented by their
principal branches, yields the correct value for the generating
integral at the reference point (1,1,1,1,1,1). If this value
changes continuously, as the parameter $p$ goes from 0 to 1, one
can be sure that the generating integral will be computed
correctly at the final point
$(\alpha_{12},\alpha_{13},\alpha_{14},\alpha_{23},\alpha_{24},\alpha_{34})$.
Therefore, continuity of this integral is a criterion of the
correct branch tracking.

Let us define a function $S(p)$ as the sum in the square brackets
of Eq.~(13) when the parameters $\{\alpha_{jk}(p)\}$ are used
instead of $\{\alpha_{jk}\}$:
\begin{equation}
S(p)=\sum_{j=1}^{4} \sum_{k=1}^{4} v(\gamma_{k}^{(j)} / \sigma) +
\sum_{j=2}^{4} u(\beta_{1}^{(1)}\beta_{1}^{(j)})~~.
\end{equation}
Our purpose is to ensure that this function is continuous along
the path from $p=0$ to $p=1$.

First, we consider a case when $\sigma^{2}(p)>0$. The function
$\sigma(p)$ is real, and all the arguments of the functions $u(z)$
and $v(z)$ in Eq.~(32) are real as well. It will be shown in
Sec.~II.E that only imaginary parts of these functions exhibit
discontinuities, when their arguments cross the branch cuts.
Because the generating integral is real, the imaginary parts of
the functions $u(z)$ and $v(z)$ in Eq.~(32) must cancel anyway.
Therefore, discontinuities in the real part of $S(p)$ may appear
near the singular points of the functions $u(z)$ and $v(z)$ only.
The singularities of different terms in Eq.~(32) should cancel one
another. However, because of possible branch changes, complete
cancellation may not happen. The formulas of Sec.~II.E suggest
that, near the singular points of $u(z)$ and $v(z)$, the real part
of the function $S(p)$ can undergo changes by $m\pi^{2}$, where
$m$ is some integer. Thus, the function $S(p)$ can have finite
discontinuities, which are integer multiples of $\pi^{2}$.

From now on, the branch tracking is only a technical problem. To
solve it, it is necessary to find all values of the parameter $p$
between 0 and 1, which correspond to singular points. They include
zeros of the sixth-order polynomial $\sigma^{2}(p)$, and values of
$p$, at which the parameters $\{\alpha_{jk}(p)\}$ satisfy any of
the conditions of Eq.~(30). Let us denote the resulting set of
numbers as $\{p_{j}\}$, $j=1...n$. The correction function, needed
to remove discontinuities of the function $S(p)$, is given by the
following expression:
\begin{equation}
C(p)=-\pi^{2} \sum_{p_{j}<p} \mathrm{Nint}
[(S(p_{j}+\epsilon)-S(p_{j}-\epsilon))/\pi^{2}]~~.
\end{equation}
Here, the function $\mathrm{Nint}[x]$ returns an integer number,
nearest to the real number $x$. The value of $\epsilon$ in actual
calculations was set to $10^{-2}$. The correct value of the
generating integral can now be determined from the formula:
\begin{equation}
I=\frac{16\pi^{3}}{\sigma} [ \sum_{j=1}^{4} \sum_{k=1}^{4}
v(\gamma_{k}^{(j)} / \sigma) + \sum_{j=2}^{4}
u(\beta_{1}^{(1)}\beta_{1}^{(j)})+C(1)\,]~~.
\end{equation}

Therefore, in the case of the real parameters $\{\alpha_{jk}\}$,
the procedure of branch tracking can be implemented without a
detailed numerical analysis of behavior of the multiple valued
functions. All we need to do is to calculate the function $S(p)$
twice for each singular point $p_{\,j}$, encountered along the
path from $p=0$ to $p=1$, and subtract discontinuities,
proportional to $\pi^{2}$. The time, needed to determine the
correction $C(1)$ in Eq.~(34), is shorter than the time, required
to carry out the recursive procedure for the family of integrals.
It does not increase the overall computation time significantly.

The case of $\sigma^{2}(p)<0$ is also straightforward. The
quantity $\sigma$ is now imaginary. The function $S(p)$ is
imaginary as well, thus giving a real value of the generating
integral. $\mathrm{Im}(S(p))$ can be expressed in terms of
Clausen's function $\mathrm{Cl}_{2}(\theta)$, which is a real
function of a real argument \cite{hil87,lev81}. Eqs.~(33) and (34)
are valid also in this case, if $S(p)$ is replaced by
$\mathrm{Im}(S(p))$, and $\sigma(p)$ is replaced by
$\mathrm{Im}(\sigma(p))$. Therefore, in both cases ($\sigma^{2}>0$
and $\sigma^{2}<0)$ the entire algorithm for analytic evaluation
of the four-particle integrals can be presented in the real form
without any use of complex numbers.

This simplified method of branch tracking has been successfully
employed in variational calculations of four-particle systems
\cite{reb96,zot98}. Therefore, the described method of branch
tracking in the case of real exponential parameters is both
theoretically correct and practically reliable.

\subsection{Branch tracking in the complex case}

Let us now describe a method of branch tracking in a general case,
when the exponential parameters, $\{\alpha_{jk}\}$, are complex
numbers. It is assumed that their real parts satisfy Eqs.~(28) and
(29). We use the same parametrization as before, but with a
complex parameter $p$\,:
\begin{equation}
\alpha_{jk}(p)=(\alpha_{jk}-1)p+1~,~~~~0 \leq \mathrm{Re}(p) \leq
1~~.
\end{equation}
As $p$ moves in the complex plane from 0 to 1, the corresponding
point in 12-dimensional space of six complex parameters moves from
(1,1,1,1,1,1) to
$(\alpha_{12},\alpha_{13},\alpha_{14},\alpha_{23},\alpha_{24},\alpha_{34})$.
The generating integral, Eq.~(13), must be a continuous function
of $p$. Moreover, its value, computed at the final point,
$\{\alpha_{jk}\}$, should not depend on a choice of the path from
$p=0$ to $p=1$. However, an optimal choice of this path can
facilitate branch tracking considerably.

Fig.~2 exhibits three examples of paths in the complex $p$ plane.
In case a), there are no singular points on or near the real axis
between 0 and 1. The path is simply a straight lime segment
between these points. In case b), there is one point, $p_{\,1}$,
at which different terms in Eq.~(13) exhibit singular behavior.
The path is the same as before, except for a small semicircle in
the vicinity of this point. In case c), there are two singular
points, $p_{\,1}$ and $p_{\,2}$, near the real axis between 0 and
1. The path is more complicated, as shown in the figure. In
general, only those singular points in the $p$ plane, which are
close to the real interval between 0 and 1, are of interest. The
path should be carefully defined in the vicinity of every such
point to allow precise analysis of behavior of all arguments of
the multiple valued functions. The values of $p$, at which
singularities may arise, can be found from the polynomial equation
$\sigma^{2}(p)=0$, and from twelve linear equations, contained in
Eq.~(30).

In order to obtain correction functions for the function $u(z)$,
defined by Eq.~(14), we have to consider behavior of this function
near its branch points 0, 1, and $\infty$:
\begin{eqnarray}
u(z \rightarrow 0)=~\frac{1}{2}\ln^{2}(-z)+u_{(0)}(z)~~;~~~
\nonumber \\
u(z \rightarrow 1)=-2\ln(z)\ln(1-z)+u_{(1)}(z)~~; \\
u(z \rightarrow \infty)=
-\frac{1}{2}\ln^{2}(-z)+u_{(\infty)}(z)~~.~~~ \nonumber
\end{eqnarray}
In these formulas, the functions with subscripts (0), (1), and
$(\infty)$ are functions, analytic in the vicinities of 0, 1, and
$\infty$, respectively.

Let us introduce the following notations. A complex function
$z(p)$ will represent any of the arguments,
$\beta_{1}^{(1)}\beta_{1}^{(j)}$, of the function $u(z)$ in
Eq.~(13). It depends on $p$ through the parameters
$\alpha_{jk}(p)$, given by Eq.~(35). Let $\{p_{j}\},~j=1...N,$
denote values of the parameter $p$, for which $z(p_{j})$ are
singular points 0, 1, $\infty$, or any points, where $z(p)$
crosses the real axis. It is assumed that $0 \leq
\mathrm{Re}(p_{j}) \leq 1$ for each $j=1...N$, and
$\mathrm{Re}(p_{j})<\mathrm{Re}(p_{j+1})$. Each point $p_{j}$ will
be characterized by an index $n_{j}$, and either integer $m_{j}$,
or real $\delta_{j}$. The index $n_{j}=1...6$ specifies a type of
singular behavior, as explained below. The number $m_{j}$ provides
information about direction, in which the real axis is crossed by
$z(p)$. We set $m_{j}=+1$, if the axis is crossed from below (i.e.
$\uparrow$), and $m_{j}=-1$, if it is crossed from above (i.e.
$\downarrow$). The real quantity $\delta_{j}$ is equal to a change
in $\mathrm{arg}(z(p)-z(p_{j}))$, when $z(p)$ moves in the
vicinity of a singular point $z(p_{j})$. If $z(p_{j})=\infty$, the
quantity $\delta_{j}$ denotes a change in $\mathrm{arg}(z(p))$.
These notations will allow us to present the algorithm of branch
tracking as a series of formulas.

\begin{figure}
\resizebox{0.9\columnwidth}{!}{\includegraphics{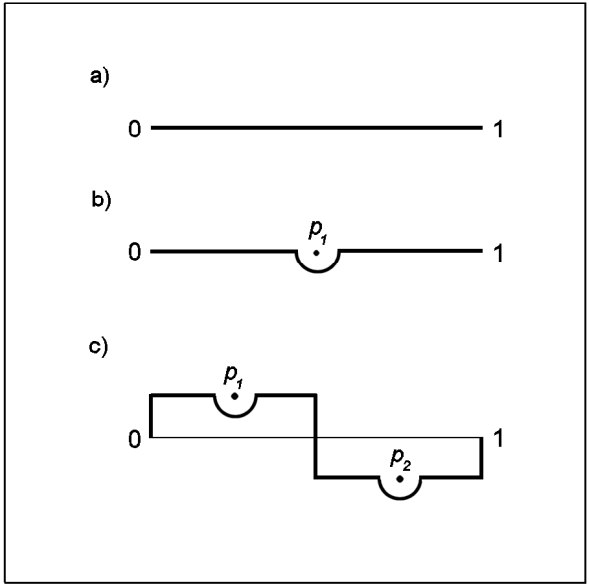}}
\caption{Examples of paths in the complex $p$ plane: a) no
singular points on or near the real axis between 0 and 1; b) one
singular point, $p_{\,1}$, on the real axis; c) two singular
points, $p_{\,1}$ and $p_{\,2}$, near the real axis. The plots are
not to scale.}
\end{figure}

Five correction functions, $u_{n_{j}}(z,j)$, are needed to restore
continuity of the computed function $u(z)$.

If $z(p)$ crosses the branch cut $]\,1,+\infty\,[$ at $z(p_{j})$,
\\ let $n_{j}=1$, and
\begin{equation}
u_{1}(z,j)=+2\pi^{2}-2m_{j} \pi\,i\, [\, \ln(-z)+U_{j} \,]~~.
\end{equation}

If $z(p)$ crosses the branch cut $]\,0,1\,[$ at $z(p_{j})$,
\\ let $n_{j}=2$, and
\begin{equation}
u_{2}(z,j)=-2\pi^{2}+2m_{j} \pi\,i\, [\, \ln(-z)+U_{j} \,]~~.
\end{equation}

If $z(p)$ moves near the singular point $z(p_{j})=1$, \\ let
$n_{j}=3$, and
\begin{equation}
u_{3}(z,j)=2 \,i\, \delta_{j} [\, \ln(z)+\hat{U}_{j} \,]~~.
\end{equation}

If $z(p)$ moves near the singular point $z(p_{j})=0$, \\ let
$n_{j}=4$, and
\begin{equation}
u_{4}(z,j)=-\delta_{j}^{2}/2 - \,i\, \delta_{j} [\, \ln(-z)+U_{j}
\,]~~.
\end{equation}

If $z(p)$ moves near the singular point $z(p_{j})=\infty$, \\ let
$n_{j}=5$, and
\begin{equation}
u_{5}(z,j)=+\delta_{j}^{2}/2 + \,i\, \delta_{j} [\, \ln(-z)+U_{j}
\,]~~.
\end{equation}

If $z(p)$ crosses the branch cut $]\,-\infty,0\,[$ of the \\
function $\ln(z)$ at $z(p_{j})$, let $n_{j}=6$.

The logarithms in these formulas are multiple valued functions
themselves. Their branches can also change, and they can exhibit
singular behavior, while an argument $z(p)$ moves further in the
complex plane. Because only the principle branch of the logarithm
is calculated by a computer, the additional terms, $U_{j}$ and
$\hat{U}_{j}$ are included to correct values of these functions.
These terms are given by the following formulas:
\begin{equation}
U_{j}=~+\sum_{\renewcommand{\arraystretch}{0.7} \begin{array}{c}
\scriptstyle k>j \\ \scriptstyle n_{k}=1,2 \end{array}}^{N}
2m_{k}\pi \,i\, ~ - \sum_{\renewcommand{\arraystretch}{0.7}
\begin{array}{c} \scriptstyle k>j \\ \scriptstyle n_{k}=4,5
\end{array}}^{N} \,i\, \delta_{k}~~~;
\end{equation}
\begin{equation}
\hat{U}_{j}=~-\sum_{\renewcommand{\arraystretch}{0.7}
\begin{array}{c} \scriptstyle k>j \\ \scriptstyle n_{k}=6 \end{array}}^{N}
2m_{k}\pi \,i\, ~ - \sum_{\renewcommand{\arraystretch}{0.7}
\begin{array}{c} \scriptstyle k>j \\ \scriptstyle n_{k}=4,5
\end{array}}^{N} \,i\, \delta_{k}~~~.
\end{equation}
The condition $n_{k}=1,2$ in these formulas means that we have to
sum up only those indices $m_{k}$, that correspond to situations,
when $z(p)$ crosses the branch cuts $]\,1,+\infty\,[$ and
$]\,0,1\,[$. The condition $n_{k}=4,5$ limits the summation of
$\delta_{k}$ to those cases, when $z(p)$ moves near the singular
points 0 and $\infty$. If $n_{k}=6$, we consider only situations,
when $z(p)$ crosses the real axes in the interval
$]\,-\infty,0\,[$.

Thus, each singular or crossing point $z(p_{j})$, encountered by
the argument $z(p)$ of the function $u(z)$, gives rise to a
correction function, $u_{n_{j}}(z,j)$, required to make $u(z)$
continuous. However, the structure of this correction function at
the end of the path, $p=1$, will depend on behavior of $z(p)$ near
all the following singular and crossing points, $z(p_{\,k}),~j < k
\leq N$. The resulting correction function, $u_{c}(z)$, obtained
after passing all the points $z(p_{j}),~j=1...N$, is given by the
following expression:
\begin{equation}
u_{c}(z)=\sum_{j=1}^{N}u_{n_{j}}(z,j)~~.
\end{equation}

In order to see, how these correction functions operate, consider
values of the principal branch of $u(z)$ at the edges of the
branch cut $]\,1,+\infty\,[$:
\begin{equation}
u(x \pm \,i\,
\epsilon)=\frac{\pi^{2}}{3}-2\mathrm{Li}_{2}(1/x)-\frac{1}{2}\ln^{2}(x)
\pm \,i\,\pi \ln(x)~~.
\end{equation}
In this formula, $x>1$ is real, and $\epsilon \rightarrow +0$.
Imagine that the branch cut is crossed from below ($\uparrow$).
Then $m_{j}=+1$, and the value of the correction function
$u_{1}(z,j)$, defined by Eq.~(37), at the point $z=x+i\epsilon$ is
equal to
\begin{displaymath}
u_{1}(x+i\epsilon,j)=-2\pi \,i\, \ln(x)~~.
\end{displaymath}
If the branch cut is crossed from above ($\downarrow$),
$m_{j}=-1$, and the value of this correction function at the point
$z=x-i\epsilon$ is equal to
\begin{displaymath}
u_{1}(x-i\epsilon,j)=+2\pi \,i\, \ln(x)~~.
\end{displaymath}
Thus, the correction function $u_{1}(z,j)$, added after the branch
cut is crossed, eliminates the finite discontinuity of the
principal branch, Eq.~(45), of the function $u(z)$ along
$]\,1,+\infty\,[$. The correction function $u_{2}(z,j)$, defined
by Eq.~(38), acts in a similar way at $]\,0,1\,[$.

Imagine now that the argument of $u(z)$ goes around the singular
point at $\infty$, starting from $z=x+i\epsilon$, and coming back
to $z=x-i\epsilon$, without crossing the branch cut along the
positive real axis. The value of $u(z)$ exhibits a singular change
(from Eq.~(45)) by
\begin{displaymath}
\Delta u = -2\pi \,i\, \ln(x)~~.
\end{displaymath}
In this case, $\delta_{j}=2\pi$, and the value of the correction
function $u_{5}(z,j)$, defined by Eq.~(41), at the point
$z=x-i\epsilon$ is equal to
\begin{displaymath}
u_{5}(x-i\epsilon,j)=+2\pi \,i\, \ln(x)~~.
\end{displaymath}
Thus, the correction function $u_{5}(z,j)$, added after $z$ has
moved near the singular point at $\infty$, eliminates the singular
contribution to the value of the function $u(z)$. The correction
functions $u_{3}(z,j)$ and $u_{4}(z,j)$, given by Eqs.~(39) and
(40), produce similar results for the other singular points.

If, in the above examples, the argument of $u(z)$ first crosses
the branch cut, and then moves around the singular point, the
correction function $u_{1}(z,j)$ has to be modified by adding
nonzero $U_{j}$ to the logarithm according to Eq.~(37).\\

The same principles of branch tracking apply to the function
$v(z)$, defined by Eq.~(15). First, we consider behavior of this
function near its branch points 1, -1, and $\infty$:
\begin{eqnarray}
v(z \rightarrow ~1)=
-\frac{1}{4}\ln^{2}[(1-z)/(1+z)]+v_{(1)}(z)~~;~~~~~~
\nonumber \\
v(z \rightarrow -1)=
~\frac{1}{4}\ln^{2}[(1+z)/(1-z)]+v_{(-1)}(z)~~;~~~~~ \\
~v(z\rightarrow \infty)
=\frac{1}{2}\ln(-z^{2}/4)\ln[(z+1)/(z-1)]+v_{(\infty)}(z)~.
\nonumber
\end{eqnarray}
In these formulas, the functions with subscripts (1), (-1), and
$(\infty)$ are functions, analytic in the vicinities of 1, -1, and
$\infty$, respectively.

Let us again consider a complex function $z(p)$, which can
represent each of the arguments $\gamma_{k}^{(j)}/\sigma$ of the
function $v(z)$ in Eq.~(13). Let $\{p_{j}\},~j=1...N$, denote
values of the parameter $p$, such that $z(p_{j})$ are singular
points 1, -1, $\infty$, or $z(p)$ crosses the real axis.  It is
assumed that their real parts form an increasing set of numbers
between 0 and 1. As before, each point $p_{j}$ is characterized by
an index $n_{j}$, and either $m_{j}$ or $\delta_{j}$. Values of
$n_{j}$ will be assigned below, and meanings of $m_{j}$ and
$\delta_{j}$ remain the same.

Five correction functions, $v_{n_{j}}(z,j)$, are used to make the
computed function $v(z)$ continuous.

If $z(p)$ crosses the branch cut $]\,1,+\infty\,[$ at $z(p_{j})$,
\\ let $n_{j}=1$, and
\begin{equation}
v_{1}(z,j)=+\pi^{2}+m_{j}\pi \,i\, \{ \, \ln[(1+z)/(1-z)]+V_{j} \,
\}~.
\end{equation}

If $z(p)$ crosses the branch cut $]-\infty,-1\,[$ at $z(p_{j})$,
\\ let $n_{j}=2$, and
\begin{equation}
v_{2}(z,j)=-\pi^{2}-m_{j}\pi \,i\, \{ \, \ln[(1+z)/(1-z)]+V_{j} \,
\}~.
\end{equation}

If $z(p)$ moves near the singular point $z(p_{j})=1$,
\\ let $n_{j}=3$, and
\begin{equation}
v_{3}(z,j)=+\frac{\delta_{j}^{2}}{4}- \,i\, \frac{\delta_{j}}{2}
\{ \, \ln[(1+z)/(1-z)]+V_{j} \, \}~.
\end{equation}

If $z(p)$ moves near the singular point $z(p_{j})=-1$,
\\ let $n_{j}=4$, and
\begin{equation}
v_{4}(z,j)=-\frac{\delta_{j}^{2}}{4}- \,i\, \frac{\delta_{j}}{2}
\{ \, \ln[(1+z)/(1-z)]+V_{j} \, \}~.
\end{equation}

If $z(p)$ moves near the singular point $z(p_{j})=\infty$,
\\ let $n_{j}=5$, and
\begin{equation}
v_{5}(z,j)=-\,i\, \delta_{j} \{ \, \ln[(z+1)/(z-1)]+\hat{V}_{j} \,
\}~.
\end{equation}

If $z(p)$ crosses the branch cut $]\,-1,1\,[$ of the \\
function $\ln[(z+1)/(z-1)]$ at $z(p_{j})$, let $n_{j}=6$.

The additional terms, $V_{j}$ and $\hat{V}_{j}$, necessary to
correct behavior of the logarithms, are given by the following
formulas:
\begin{equation}
V_{j}=-\sum_{\renewcommand{\arraystretch}{0.7}
\begin{array}{c} \scriptstyle k>j
\\ \scriptstyle n_{k}=1,2 \end{array}}^{N} 2m_{k}\pi \,i\, +
\sum_{\renewcommand{\arraystretch}{0.7} \begin{array}{c}
\scriptstyle k>j \\ \scriptstyle n_{k}=3
\end{array}}^{N} \,i\, \delta_{k} -
\sum_{\renewcommand{\arraystretch}{0.7} \begin{array}{c}
\scriptstyle k>j \\ \scriptstyle n_{k}=4
\end{array}}^{N} \,i\, \delta_{k}~;
\end{equation}
\begin{equation}
\hat{V}_{j}=+\sum_{\renewcommand{\arraystretch}{0.7}
\begin{array}{c} \scriptstyle k>j
\\ \scriptstyle n_{k}=6 \end{array}}^{N} 2m_{k}\pi \,i\, +
\sum_{\renewcommand{\arraystretch}{0.7} \begin{array}{c}
\scriptstyle k>j \\ \scriptstyle n_{k}=3
\end{array}}^{N} \,i\, \delta_{k} -
\sum_{\renewcommand{\arraystretch}{0.7} \begin{array}{c}
\scriptstyle k>j
\\ \scriptstyle n_{k}=4
\end{array}}^{N} \,i\, \delta_{k}~.
\end{equation}

As in the previous case, a correction function $v_{n_{j}}(z,j)$
has to be added to the function $v(z)$ every time its argument
$z(p)$ passes a singular or crossing point $z(p_{j})$. This way,
the calculated function $v(z)$ can be made continuous. However,
the form of this correction function at the end of the path
depends on behavior of $z(p)$ near all the points $z(p_{\,k})$,
following $z(p_{j})$. The resulting correction function,
$v_{c}(z)$, is the following:
\begin{equation}
v_{c}(z)=\sum_{j=1}^{N}v_{n_{j}}(z,j)~~.
\end{equation}

Let us now briefly discuss the effect of using these correction
functions. Consider values of $v(z)$ at the edges of the branch
cut $]\,1,+\infty\,[$:
\begin{eqnarray}
v(x \pm i\epsilon)=\frac{1}{2}\mathrm{Li}_{2}[2/(1+x)]-
\frac{1}{2}\mathrm{Li}_{2}[2/(1-x)]+~ \nonumber \\
+\frac{1}{2}\ln^{2}[2/(1+x)]-\frac{1}{2}\ln^{2}[2/(x-1)]\pm~\\
\pm \,i\, \frac{\pi}{2} \ln[(x-1)/(x+1)]~~.~~~~~~~~~~~~~~~~
\nonumber
\end{eqnarray}
Here, $x>1$ is real, and $\epsilon \rightarrow +0$. Imagine that
the branch cut is crossed from below ($\uparrow$). Then
$m_{j}=+1$, and a value of the correction function, $v_{1}(z,j)$,
defined by Eq.~(47), at a point $z=x+i\epsilon$ is equal to
\begin{displaymath}
v_{1}(x+i\epsilon,j)=-\,i\,\pi\ln[(x-1)/(x+1)]~~.
\end{displaymath}
If the branch cut is crossed from above ($\downarrow$), then
$m_{j}=-1$, and a value of this correction function at $z=x-i
\epsilon$ is equal to
\begin{displaymath}
v_{1}(x-i\epsilon,j)=+\,i\,\pi\ln[(x-1)/(x+1)]~~.
\end{displaymath}
Therefore, the function $v_{1}(z,j)$, added to the function $v(z)$
after the branch cut is crossed, removes the discontinuity of the
principal branch along $]\,1,+\infty\,[$. The correction function
$v_{2}(z,j)$, given by Eq.~(48), makes $v(z)$ continuous at
$]-\infty,-1\,[$.

Imagine now that the argument of $v(z)$ moves around the singular
point +1, starting from $z=x+i\epsilon$ and returning to
$z=x-i\epsilon$, without crossing the branch cut. The value of
$v(z)$ undergoes a change (from Eq.~(55)) by
\begin{displaymath}
\Delta v=-\,i\,\pi \ln[(x-1)/(x+1)]~~.
\end{displaymath}
Because $\delta_{j}=2\pi$, a value of the correction function
$v_{3}(z,j)$, defined by Eq.~(49), at $z=x-i\epsilon$ is equal to
\begin{displaymath}
v_{3}(x-i\epsilon,j)=+\,i\,\pi\ln[(x-1)/(x+1)]~~.
\end{displaymath}
Thus, by adding the correction function $v_{3}(z,j)$, it is
possible to eliminate the singular contribution to the value of
$v(z)$, when $z$ goes around the singular point +1. The correction
functions $v_{4}(z,j)$ and $v_{5}(z,j)$, given by Eqs.~(50) and
(51), produce the same results for the other two singular points.

If, in the above examples, the argument of $v(z)$ first crosses
the branch cut, and then moves around the singular point, the
correction function $v_{1}(z,j)$ should be modified by adding
nonzero $V_{j}$ according to Eq.~(47).\\

It is important to note that the function $\sigma(z)$ is also a
multiple valued function. Its principal branch, defined by
Eq.~(27), changes sign each time the argument $z$ crosses the
branch cut along the positive real axis. If this happens $N$ times
while the parameter $p$ changes from 0 to 1, the corrected value
$\sigma_{c}(z)$ of this function at $p=1$ is equal to
\begin{equation}
\sigma_{c}(z)=(-1)^{N}\sigma(z)~~,
\end{equation}
where $\sigma(z)$ is the value of the principal branch of the
complex square root.

We are now in a position to write a corrected expression for the
generating integral:
\begin{equation}
I=\frac{16\pi^{3}}{\sigma_{c}} [ \sum_{j=1}^{4} \sum_{k=1}^{4}
(v+v_{c})(\gamma_{k}^{(j)} \! / \sigma_{c}) + \sum_{j=2}^{4}
(u+u_{c})(\beta_{1}^{(1)}\beta_{1}^{(j)}) ]
\end{equation}
In this formula, each of 16 terms with the function $v(z)$
contains its own correction function $v_{c}(z)$, which is
characterized by its own $N$ and sets of numbers $\{p_{\,l}\}$,
$\{n_{l}\}$, $\{m_{l}\}$, and $\{\delta_{l}\}$. The same is true
for each of the three terms with the function $u(z)$.

Eq.~(57) is profoundly different from the original formula,
Eq.~(13), for the generating integral. In Eq.~(13), the functions
$u(z)$, $v(z)$, and $\sigma(z)$ are expressed in terms of the
multiple valued logarithm and square root. When Eq.~(57) is used,
it is assumed, on the contrary, that all the logarithms and square
roots are represented by their principal branches, and, therefore,
can be readily evaluated by a computer. The multiple valued nature
of the functions $u(z)$, $v(z)$, and $\sigma(z)$ is taken into
account explicitly by means of the additional correction terms and
factors. Also, singular contributions from different terms in
Eq.~(13) are expected to cancel each other to yield a correct
value for the generating integral, Eq.~(11). When we use Eq.~(57),
all the singular contributions from different terms are cancelled
explicitly and separately, so that each function
$(v+v_{c})(\gamma_{k}^{(j)} / \sigma_{c})$ or
$(u+u_{c})(\beta_{1}^{(1)}\beta_{1}^{(j)})$ is continuous along
the path from (1,1,1,1,1,1) to
$(\alpha_{12},\alpha_{13},\alpha_{14},\alpha_{23},\alpha_{24},\alpha_{34})$.
As a result, the generating integral, obtained from Eq.~(57), is a
continuous function of the complex parameters $\{\alpha_{jk}\}$.
Thus, the problem of branch tracking is successfully solved in the
most general case.

Let us consider an example of branch tracking for the function
$v(z)$ with the argument $z(p)$. A sample path in the complex $p$
plane is displayed in Fig.~3a, and the corresponding path from
$z(0)=A$ to $z(1)=B$ in the complex $z$ plane is shown in Fig.~3b.
There are five points of interest: $z(p_{\,1})=1$,
$z(p_{\,3})=\infty$, and $z(p_{\,5})=-1$ are singular points of
the function $v(z)$; $z(p_{\,2})$ and $z(p_{\,4})$ are points,
where the argument $z(p)$ crosses the branch cuts. According to
the chosen classification: $n_{1}=3,~\delta_{1}=\pi$;
$n_{2}=1,~m_{2}=+1$; $n_{3}=5,~\delta_{3}=\pi$;
$n_{4}=2,~m_{4}=-1$; $n_{5}=4,~\delta_{5}=\pi$. Using
Eqs.~(47)-(53), one can easily obtain expressions for the
correction functions:
\begin{eqnarray}
v_{3}(z,1)=~\frac{\pi^{2}}{4}-\,i\,\frac{\pi}{2} \{ \,
\ln[(1+z)/(1-z)]-\,i\,\pi \, \}~~;
\nonumber \\
v_{1}(z,2)=~~\pi^{2}+\,i\, \pi \{ \, \ln[(1+z)/(1-z)]+\,i\,\pi \,
\} ~~; \,
\nonumber \\
v_{5}(z,3)=~~0~~-\,i\, \pi \{ \, \ln[(z+1)/(z-1)]-\,i\,\pi \,
\}~~;
\nonumber \\
v_{2}(z,4)=-\pi^{2}+\,i\, \pi \{ \, \ln[(1+z)/(1-z)]-\,i\,\pi \,
\} ~~;
\nonumber \\
v_{4}(z,5)=-\frac{\pi^{2}}{4}-\,i\,\frac{\pi}{2} \{\,
\ln[(1+z)/(1-z)]+~0\, \}~~. \nonumber
\end{eqnarray}
The resulting correction function, $v_{c}(z)$, is
\begin{eqnarray}
v_{c}(z)=-\frac{3\pi^{2}}{2}+\,i\, \pi \ln[(1+z)/(1-z)]- \nonumber
\\ -\,i\, \pi \ln[(z+1)/(z-1)]~~.~ \nonumber
\end{eqnarray}
One can see that this function is different from zero, even if
$A=B$, i.e. the contour is closed. This is not surprising. Even
though $v(z)$ is represented by its principal branch, the sum
$v(z)+v_{c}(z)$ is still a multiple valued function. Its value
generally undergoes a finite change, if its argument traverses a
closed loop, encircling branch points. Consider a value of
$v_{c}(z)$ at $z=x-i\epsilon$, where $-1<x<1$, and $\epsilon
\rightarrow +0$. In this case, the logarithms cancel, and
\begin{displaymath}
v_{c}(x-i\epsilon)=-\pi^{2}/2~~.
\end{displaymath}
Contributions of this type from different terms in Eq.~(57)
produce an additional constant $m\pi^{2}$, needed to correct a
value of the generating integral in the case of real parameters
$\{\alpha_{jk}\}$. Thus, Eq.~(34) is a particular case of
Eq.~(57).

\begin{figure}
\resizebox{0.9\columnwidth}{!}{\includegraphics{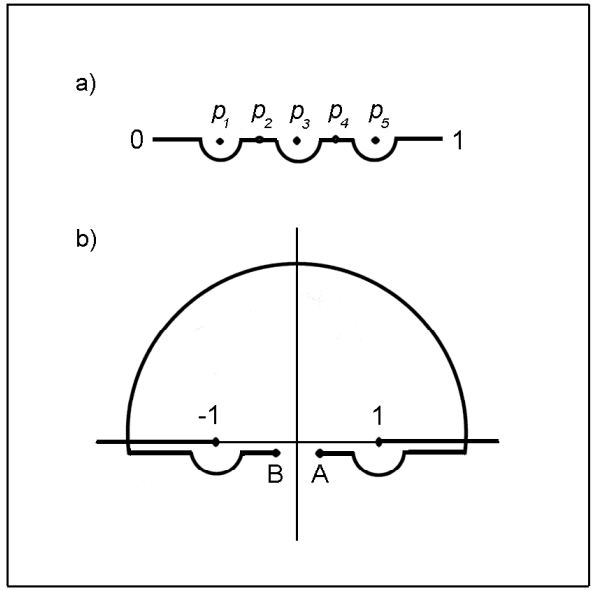}}
\caption{Illustration of branch tracking in the complex case: a)
the path in the complex $p$ plane, chosen to avoid singular
points, $p_{\,1}$, $p_{\,3}$, and $p_{\,5}$; b) the corresponding
path in the complex $z$ plane from $z(0)=A$ to $z(1)=B$, where
$z(p)$ is an argument of the multiple valued function $v(z)$.}
\end{figure}

This example demonstrates that the branch tracking in the general
case requires a comprehensive numerical analysis of behavior of
all the arguments in Eq.~(13).

\subsection{Numerical procedure and results}

Practical implementation of the method, described in the previous
sections, is inevitably a very complicated task. Detailed
information about the recursive procedure, needed to compute the
family of integrals, Eq.~(10), can be found in \cite{hil87}. Here
we describe only the procedure for numerical branch tracking.

First, the set of points, $\{p_{\,k}\}$, at which different terms
in Eq.~(13) can exhibit singular behavior, is determined. This is
done by solving the sixth-order equation $\sigma^{2}=0$, and
linear equations of Eq.~(30), with the parametrization according
to Eq.~(35). Only those values of $p$, that lie in or near the
real interval $]\,0,1\,[$, are included in the set $\{p_{\,k}\}$.
Then, a path from 0 to 1 in the complex $p$ plane is chosen.
Fig.~2 gives an idea of this. The whole path is shifted downward
by a small imaginary quantity $i\epsilon$ to avoid possible
ambiguities, when arguments of the functions $u(z)$ and $v(z)$ are
real. All the arguments in Eq.~(57) are computed at the final
point of this path, $p=1-i\epsilon$. In actual calculations,
$\epsilon$ was set to $10^{-28}$. This did not affect values of
the integrals, but was enough to shift the arguments from the real
axis.

The path in the complex $p$ plane is divided into small intervals,
as shown in Fig.~4. The intervals $]\,P_{\,l},P_{\,l+1}\,[$, into
which the linear segments between the singular points,
$\{p_{\,k}\}$, are divided, have a typical length of $10^{-2}$.
Each small semicircle beneath a singular point, $p_{\,k}$, has a
radius $r=10^{-9}$, and divided into six parts. The corresponding
boundary points are
\begin{equation}
P_{\,l}=p_{\,k}+r\exp[\,i\,(\pi\,l/6)\,]-\,i\,\epsilon~,~~~l=0...6.
\end{equation}

\begin{figure}
\resizebox{0.8\columnwidth}{!}{\includegraphics{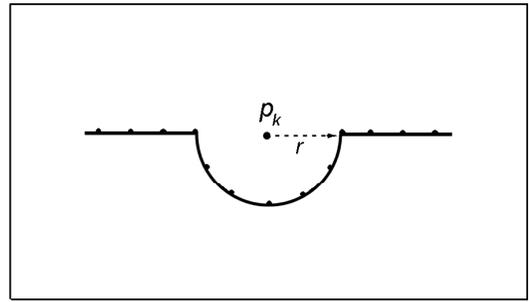}}
\caption{Segmentation of the path in the complex $p$ plane, needed
to analyze behavior of arguments of the multiple valued functions
numerically.}
\end{figure}

In order to obtain full information about behavior of different
arguments in Eq.~(13), all these arguments should be computed at
all the points $P_{\,l}$ along the path. The quantities
$\gamma_{k}^{(j)}$, $\beta_{k}^{(j)}$, and $\sigma^{2}$, given by
Eqs.~(16)-(20), are simple functions of $\{\alpha_{jk}(p)\}$, so
this calculation can be performed almost immediately. The values
of each argument are analyzed, and the numbers $N$, $n_{j}$,
$m_{j}$, and $\delta_{j}$, $j=1...N$, needed to apply the formulas
of Sec.~II.E, are determined. This procedure works as follows. To
find out, if an argument crosses the real axis, the imaginary
parts of its values, computed at points $P_{\,l}$ and $P_{\,l+1}$,
are compared. If they have opposite signs, dichotomy is used to
reduce the interval, and determine, where the real axis is
crossed, and in which direction. This is also done for the
intervals along each small semicircle, but without the dichotomy.
In this manner, all the crossing points can be found and analyzed.
At each interval, $]\,P_{\,l},P_{\,l+1}\,[$, this analysis is
carried out for $\sigma^{2}$ first. If $\sigma^{2}$ crosses the
positive real axis, when $p$ is between $P_{\,l}$ and $P_{\,l+1}$,
the sign of $\sigma$ should be changed. Then, all
$\gamma_{k}^{(j)}$, computed at $P_{\,l+1}$, should be multiplied
by (-1), and all $\beta_{k}^{(j)}$ -- inverted. The quantity
$\delta_{j}$ is determined as follows:
\begin{equation}
\delta_{j}=\sum_{l=0}^{5}[\,
\mathrm{arg}(z(P_{\,l+1}))-\mathrm{arg}(z(P_{\,l})) \,]~~.
\end{equation}
Here, the points $P_{\,l}$ are specified by Eq.~(58), and $z$ can
stand for any of the arguments $\gamma_{k}^{(j)} / \sigma$ and
$\beta_{1}^{(1)}\beta_{1}^{(j)}$ as functions of
$\{\alpha_{jk}(p)\}$. The right hand side of this formula is
presented as a sum, because as $p$ changes from $P_{\,0}$ to
$P_{\,6}$, the argument $z(p)$ may go around the singular point,
$z(p_{\,k})$ several times. Thus, behavior of each argument in the
vicinity of each singular point can be analyzed. The described
procedure of numerical branch tracking provides all the
information, necessary for successful use of Eq.~(57).

The entire algorithm for analytic evaluation of the four-particle
integrals with complex parameters was tested in four different
ways.

First, real parameters $\{\alpha_{jk}\}$ were used, and the
results, obtained using the complex algorithm of Sec.~II.E, were
compared with results, provided by the method of Sec.~II.D for the
real case. The real parts of the computed integrals, Eq.~(10),
were invariably in excellent agreement. The imaginary parts, given
by the new method, were at least 20 orders of magnitude smaller
than the real parts, and could be considered negligible.
Therefore, the complex algorithm works correctly for any
acceptable real parameters.

Second, the parameters $\{\alpha_{jk}\}$ were multiplied by an
arbitrarily chosen complex number $\lambda$. A resulting integral,
Eq.~(10), with a particular set $\{n_{jk}\}$ must be equal to its
original value, multiplied by $f=-1/(-\lambda)^{K}$, where
$K=n_{12}+n_{13}+n_{14}+n_{23}+n_{24}+n_{34}+3$. Various
$\lambda$'s were used, and values of the integrals, calculated
directly, were compared with the rescaled original values.
Remarkable agreement was observed in all these cases. Note that
different values of $\lambda$ correspond to different paths in the
space of parameters according to Eq.~(35).

Third, if two exponential parameters, $\alpha_{13}$ and
$\alpha_{24}$, are equal to zero, the six Coulomb integrals and
one overlap integral, needed to determine matrix elements of the
Hamiltonian according to Eqs.~(4)-(9), can be obtained
analytically in terms of rational functions and logarithms. Values
of these integrals, calculated with various sets of complex
parameters, $\alpha_{12}$, $\alpha_{14}$, $\alpha_{23}$,
$\alpha_{34}$, were compared with the same integrals, computed
using the new method. They were always in complete agreement.

Fourth, different paths in the complex $p$ plane were chosen. They
included singularities, located not only near the real axis, but
also further away. The results did not depend on the choice of the
path. This fact suggests that the described method of numerical
branch tracking is stable and reliable. Of course, the path in
actual calculations should be as simple as possible, provided that
all nearby singularities are carefully taken into account.

Table~I displays values of the integrals for three different sets
of parameters $\{\alpha_{jk}\}$, used to test the computer
program. Many other sets of parameters were also considered. All
the integrals were calculated using the general algorithm for
numerical branch tracking, described in Sec.~II.E. Implementation
of this algorithm requires quadruple precision. The program
computes a family of 64 integrals, Eq.~(10), with two possible
values for every index: $n_{jk}=0,1$. Only seven integrals,
necessary to obtain matrix elements of the Hamiltonian according
to Eqs.~(4)-(9), are presented in Table~I for each set of
parameters.

\begin{table*}
\caption{Examples of four-particle integrals evaluated using the
algorithm for numerical branch tracking in the complex case.}
\begin{ruledtabular}
\begin{tabular}{lccc}
$\{\,\alpha_{jk}\}$  &  $\{\,n_{jk}\}$  &  Re\,($J$\,)  &  Im\,($J$\,) \\
\hline
$\alpha_{12}=~~1.56~~~~~~~$ & 011111 & ~0.20550889174003868108D+03 & -0.52323042463487687803D\,-\,21\\
$\alpha_{13}=-0.69~~~~~~~$ & 101111 & ~0.49701602825033406834D+02 & ~0.37462639365280834442D\,-\,21\\
$\alpha_{14}=~~2.71~~~~~~~$ & 110111 & ~0.33278606420131925558D+03 & ~0.29262335382214159490D\,-\,21\\
$\alpha_{23}=~~1.75~~~~~~~$ & 111011 & ~0.69156207020089168507D+02 & ~0.39044885805975934492D\,-\,21\\
$\alpha_{24}=~~1.42~~~~~~~$ & 111101 & ~0.20401147280211976836D+03 & ~0.44225408142944401598D\,-\,21\\
$\alpha_{34}=-0.50~~~~~~~$ & 111110 & ~0.50046583501959809463D+02 & ~0.75995078405826138804D\,-\,21\\
~~~~~~~~~~~~~~~~~~~~~~~~~~ & 111111 & ~0.21644781505854395857D+03 & ~0.32039794574654543500D\,-\,20\\
\hline
$\alpha_{12}=~~1.56*(1+0.5\,i)$ & 011111 &-0.70977575942226172269D+02 & ~0.45253255249692588012D+02\\
$\alpha_{12}=-0.69*(1+0.5\,i)$ & 101111 &-0.17165677159247121876D+02 & ~0.10944340655611068217D+02\\
$\alpha_{12}=~~2.71*(1+0.5\,i)$ & 110111 &-0.11493589374343266153D+03 & ~0.73279810811752133344D+02\\
$\alpha_{12}=~~1.75*(1+0.5\,i)$ & 111011 &-0.23884805635825330948D+02 & ~0.15228263175782374191D+02\\
$\alpha_{12}=~~1.42*(1+0.5\,i)$ & 111101 &-0.70460405295819730405D+02 & ~0.44923522162040663029D+02\\
$\alpha_{12}=-0.50*(1+0.5\,i)$ & 111110 &-0.17284824763945988644D+02 & ~0.11020305731851711925D+02\\
~~~~~~~~~~~~~~~~~~~~~~~~~~~~~~~ & 111111 &-0.40739676150047588287D+02 & ~0.68031854740817630022D+02\\
\hline
$\alpha_{12}=1.29+1.19\,i$ & 011111 & ~0.41299141847575234393D+01 & -0.13354699318829522025D+01\\
$\alpha_{13}=0~~~~~~~~~~~$ & 101111 & ~0.25568585205838373519D+01 & -0.14420903111112389762D+01\\
$\alpha_{14}=2.53-1.32\,i$ & 110111 & ~0.39327459787814931363D+01 & -0.64522069912295906967D+01\\
$\alpha_{23}=1.86+1.44\,i$ & 111011 & ~0.56761553854820761165D+01 & -0.12198339708832004631D+01\\
$\alpha_{24}=0~~~~~~~~~~~$ & 111101 & ~0.38294186820466745046D+01 & -0.25317810705579310712D+01\\
$\alpha_{34}=0.65-0.93\,i$ & 111110 & ~0.26044120278339787630D+01 & -0.23151954768733032766D+01\\
~~~~~~~~~~~~~~~~~~~~~~~~~~ & 111111 & ~0.37849264531713841033D+01 & -0.28372153117607227596D+01\\
\end{tabular}
\end{ruledtabular}
\end{table*}

Our results demonstrate that the developed algorithm allows
precise evaluation of the four-particle integrals with arbitrary
complex parameters, provided that the integrals themselves
converge.

The described method makes it possible to use the highly versatile
exponential-trigonometric basis functions in variational
calculations of four-particle Coulomb systems. In order to
illustrate efficiency of the new basis, we would like to mention
some results, obtained previously \cite{zot00} for the following
systems: $e^{+}e^{-}e^{+}e^{-}$, $p^{+}\mu^{-}p^{+}\mu^{-}$,
$\mu^{+}e^{-}\mu^{+}e^{-}$, and $p^{+}e^{-}p^{+}e^{-}$. The
calculations were performed using one exponential-trigonometric
basis function:
\begin{equation}
\Psi=\hat{S}
\exp(-\sum_{j<k}^{4}A_{jk}r_{jk})\sin(\sum_{j<k}^{4}B_{jk}r_{jk}+C)~~.
\end{equation}
This function includes 12 real nonlinear parameters, $\{A_{jk}\}$
and $\{B_{jk}\}$, and one linear parameter, $\tan(C)$. It can be
considered a linear combination of two exponential functions,
Eq.~(2), with the complex parameters $A_{jk} \pm \,i\, B_{jk}$.
The operator $\hat{S}$ ensures that this function has correct
symmetry with respect to permutations of particles.

All integrals, necessary to determine matrix elements of the
Hamiltonian, Eq.~(1), with the function $\Psi$, were computed
according to the method, described in this paper. The nonlinear
parameters were subjected to careful gradient optimization. For
more details about this calculation, see \cite{zot00}.

Table~II exhibits values of the ground-state energy, $E$, for
$e^{+}e^{-}e^{+}e^{-}$, $p^{+}\mu^{-}p^{+}\mu^{-}$,
$\mu^{+}e^{-}\mu^{+}e^{-}$, and $p^{+}e^{-}p^{+}e^{-}$, determined
using the variational method with the trial function $\Psi$. The
table also displays the most accurate energy values, $E_{0}$,
available for these systems \cite{usu98,fro97,reb97,kin99}. One
can see that the relative errors are 0.2\%, 0.7\%, 2.4\%, and
3.6\%, respectively. The results for two adiabatic systems,
$\mu^{+}e^{-}\mu^{+}e^{-}$ and $p^{+}e^{-}p^{+}e^{-}$, with very
low mass ratios, $m/M$, are very impressive. Neither Gaussian, nor
exponential functions are even nearly as efficient \cite{zot00}.
Thus, a single symmetrized exponential-trigonometric basis
function, Eq.~(60), provides a remarkable accuracy in variational
calculations of various four-particle systems.

\begin{table}
\caption{Ground-state energy, $E$, of four molecules, computed
with a single exponential-trigonometric basis function. The most
accurate values, $E_{0}$, of this energy are taken from Refs.~1,
2, 14 and 15, respectively. Atomic units.}
\begin{ruledtabular}
\begin{tabular}{ccccc}
          System           &   $m/M$   &   $E$    &  $E_{0}$ & Error \\
\hline
$e^{+}e^{-}e^{+}e^{-}$     & 1         & -0.514956 & -0.516003 & 0.2\% \\
$p^{+}\mu^{-}p^{+}\mu^{-}$ & 0.1126095 & -198.2056 & -199.6294 & 0.7\% \\
$\mu^{+}e^{-}\mu^{+}e^{-}$ & 0.0048363 & -1.113198 & -1.141000 & 2.4\% \\
$p^{+}e^{-}p^{+}e^{-}$     & 0.0005446 & -1.122378 & -1.164025 & 3.6\% \\
\end{tabular}
\end{ruledtabular}
\end{table}

\section{Conclusion}

The method for analytic evaluation of four-particle integrals with
complex parameters, described in this paper, can be regarded as
both further theoretical development and practical implementation
of the original method by Fromm and Hill \cite{hil87}. Validity of
this method is not limited to the case of real parameters.
Moreover, because the integrals are expressed in terms of multiple
valued complex functions, it is more natural to consider a general
case when all the parameters are complex. The original formula,
Eq.~(13), for the generating integral can be used only in the
immediate vicinity of the standard reference point where all the
parameters are equal to 1. The procedure of numerical branch
tracking, proposed in this paper, allows computation of the
integrals at any other point in the space of six complex
parameters, by taking into account all branch changes along the
path. The simplified method of branch tracking for real parameters
is also discussed.

The new method makes possible high precision variational solution
of the Coulomb four-body problem in the basis of
exponential-trigonometric functions. The first calculations have
shown high efficiency of this basis \cite{zot00}. They have also
demonstrated correctness of the branch tracking algorithm,
described in this paper.

However, if the full potential of the exponential-trigonometric
basis is to be revealed, an efficient procedure for selecting
optimal values of the nonlinear parameters is necessary. Ideally,
all the parameters should be chosen \emph{a priori}, and all
matrix elements -- computed only once. Such a procedure has been
developed by the authors for the case of adiabatic three-particle
systems \cite{zot94}. All nonlinear parameters of the
exponential-trigonometric functions had been chosen \emph{before}
the computation, which yielded 10 correct significant figures for
the ground state energy of $H_{2}^{+}$ \cite{zot94}. We believe
that the exponential-trigonometric basis can provide similar
precision in calculations of four-particle
systems.\\


\begin{thebibliography}{99}

\bibitem{usu98} J.~Usukura, K.~Varga, and Y.~Suzuki, Phys. Rev. A
\textbf{58}, 1918 (1998); Y.~Suzuki and J.~Usukura, Nucl. Instr.
Meth. Phys. Res. B \textbf{171}, 67 (2000).
%
\bibitem{fro97} A.~M.~Frolov and V.~H.~Smith, Phys. Rev. A
\textbf{55}, 2435 (1997); J. Phys. B \textbf{29}, L433 (1996).
%
\bibitem{suz98} Y.~Suzuki, K.~Varga, and J.~Usukura, Nucl. Phys. A
\textbf{631}, 91 (1998); K.~Varga, Nucl. Phys. A \textbf{684}, 209
(2001).
%
\bibitem{hil87} D.~M.~Fromm and R.~N.~Hill, Phys. Rev. A
\textbf{36}, 1013 (1987).
%
\bibitem{lev81} L.~Lewin, \emph{Polylogarithms and Associated
Functions} (North-Holland, Amsterdam, 1981).
%
\bibitem{reb96} T.~K.~Rebane, V.~S.~Zotev, and O.~N.~Yusupov, Zh.
Eksp. Teor. Fiz. \textbf{110}, 55 (1996) [JETP \textbf{83}, 28
(1996)].
%
\bibitem{zot98} V.~S.~Zotev and T.~K.~Rebane, Opt. Spektrosk.
\textbf{85}, 935 (1998) [Opt. Spectrosk. \textbf{85}, 856 (1998)].
%
\bibitem{fro01} The exponential basis for variational solution of
the four-body problem was discussed recently by A.~M.~Frolov and
V.~H.~Smith, J. Chem. Phys. \textbf{115}, 1187 (2001). They,
however, were unable to compute any four-particle integrals.
%
\bibitem{reb90} T.~K.~Rebane and O.~N.~Yusupov, Zh. Eksp. Teor.
Fiz. \textbf{98}, 1870 (1990) [JETP \textbf{71}, 1050 (1990)].
%
\bibitem{zot94} V.~S.~Zotev and T.~K.~Rebane, Opt. Spektrosk.
\textbf{77}, 733 (1994) [Opt. Spectrosk. \textbf{77}, 654 (1994)].
%
\bibitem{zot00} V.~S.~Zotev and T.~K.~Rebane, Yad. Fiz.
\textbf{63}, 46 (2000) [Phys. At. Nucl. \textbf{63}, 40 (2000)].
%
\bibitem{reb93} T.~K.~Rebane, Opt. Spektrosk. \textbf{75}, 945
(1993) [Opt. Spectrosk. \textbf{75}, 557 (1993)].
%
\bibitem{reb99} T.~K.~Rebane, Opt. Spektrosk. \textbf{86}, 33
(1999) [Opt. Spectrosk. \textbf{86}, 26 (1999)].
%
\bibitem{reb97} T.~K.~Rebane, Yad. Fiz. \textbf{60}, 1628 (1997)
[Phys. At. Nucl. \textbf{60}, 1483 (1997)].
%
\bibitem{kin99} D.~B.~Kinghorn and L.~Adamowitz, Phys. Rev. Lett.
\textbf{83}, 2541 (1999).
%
\end{thebibliography}
\end{document}